# Evaluation of A National Digitally-Enabled Health Promotion Campaign for Mental Health Awareness using Social Media Platforms Tik Tok, Facebook, Instagram, and YouTube

***Samantha Bei Yi Yan*[1*] *and Dinesh Visva Gunasekeran*[1-3*]**, *Caitlyn Tan[2], Kai En Chan[2], Caleb Tan[1], Charmaine Shi Min Lim[2], Audrey Chia[4], Hsien-Hsien Lei[5],* ***Robert Morris*[1,2#] *and Janice Huiqin Weng*[1#]** *and the MINDLINE Study Group*

*These two authors contributed equally to this work and share ***first authorship***.
[#]These two authors contributed equally to this work and share ***senior authorship (Co-last author)***

**Author Affiliations:**
[1]MOH Office for Healthcare Transformation, Singapore
[2]Yong Loo Lin School of Medicine, National University of Singapore, Singapore
[3]Ophthalmology Academic Clinical Program (ACP), Duke-NUS Medical School, Singapore
[4]School of Business, National University of Singapore, Singapore
[5]NUS Saw Swee Hock School of Public Health (NUS-SSHSPH), Singapore, Singapore

***MINDLINE Study Group***: *Samantha Bei Yi Yan[1], Dinesh Visva Gunasekeran[1-3], Caitlyn Tan[2], Kai En Chan[2], Caleb Tan[1], Charmaine Shi Min Lim[2], Audrey Chia[4], Hsien-Hsien Lei[5], Onno P Kampman[1], Creighton Heaukulani[1], Yan Yan Hu[1], Julian Kui Yu Chang[1], Akash Perera[1], Ye Sheng Pang[1], Alton Ming Kai Chew[2], Krishna Vikneson[2], Kishanti Ponampalam[2], Kang-An Wong[2], Kavita Govintharasah[2], Kayshandra Tangasamy[2], Grace Enyan Aik[2], Pavanni Ponampalam[2], Hazirah Hoosainsah[1], Charmaine Ruling Lim[1], Thisum Kankanamge Thisum[1], XinYi Hong[1], Mary Grace Yeo[1], Robert Morris[1,2], Janice Huiqin Weng[1].*

***Corresponding Author:***
Asst. Prof. Dr. Dinesh Visva Gunasekeran (mdcdvg@nus.edu.sg)
Assistant Professor, Yong Loo Lin School of Medicine, National University of Singapore (NUS), Singapore
Assistant Professor, Ophthalmology Academic Clinical Program (ACP), Duke-NUS Medical School, Singapore
Clinician Scientist (AI & Digital Health), MOH Office of Healthcare Transformation (MOHT), Singapore

*Abstract*— **Mental health disorders rank among the leading contributors to the global burden of diseases and often start in adolescence, making it a crucial period for health promotion. This study was conducted to evaluate the effectiveness of digitally-enabled mental health promotion (DEHP) campaigns reported globally, and describe our experience with a national multi-platform DEHP campaign in Singapore distributed across social media platforms YouTube, Facebook, Instagram, and TikTok. Campaign materials were developed using a marketing funnel framework targeted at promoting symptom awareness, coping strategies, and/or patient navigation to mindline.sg, a national digital health platform, as the intended endpoint for user engagement and support. The campaign generated 3.49 million impressions and reached an estimated 1.39 million unique Singapore residents (>20% of the resident population). Our results are compared against the strategies and results of other published DEHP campaigns. This study demonstrates that DEHP campaigns can effectively achieve national engagement for mental health awareness. Keywords— Health Promotion, Digital Health, Social Media, Public Health, Mental Health.**

## I. INTRODUCTION

Mental health disorders represent a significant and growing global burden, contributing to a substantial portion of Years Lived with Disability (YLD) across all ages[1]. Affecting over 10% of youth globally, these disorders often start in adolescence, making it a crucial period for mental health-related health promotion[2]. Despite this detrimental impact, social stigma and logistical barriers continue to impede timely diagnosis and intervention[3]. The rapid proliferation of digital technologies has expanded opportunities for health promotion, particularly through social media and other digital platforms[4].

Platforms like YouTube, Facebook, Instagram, and TikTok now host a substantial volume of user generated content (UGC) consumed by young adult and working-age users—populations at higher risk of mental health challenges yet often underserved by traditional health promotion outreach. In response, modern health promotion methodologies that are being applied in the public health and community care fields increasingly leverage digital and social marketing strategies to reach broad audiences and deliver tailored content to specific population segments[5,6]. Recent literature has highlighted the efficacy of these campaigns and the importance of creative, user-centred design and multi-channel outreach for effective health promotion[7,8].

However, social media platforms present a potential double-edged sword for public health. Its potential negative impact on public health includes the dissemination of misinformation or fear-inducing information. On the other hand, it also has potential for positive impact through the dissemination of verified public health information, especially when content is designed to be relatable, engaging, and emotionally resonant[4,9]. These platforms can support patient education, public health campaigns, and behaviour change interventions, with mass media campaigns demonstrating increased effectiveness when reinforced across multiple dissemination channels[8,10]. Therefore, social media can offer scalable and interactive means to disseminate health information to large intended audiences[4,11].

In line with these developments, governments and healthcare organizations now deploy narrative videos, social media posts, and curated ads to raise awareness of health issues[5,12,13]. In Singapore, mental health is a national priority amid persistent stigma and limited specialist resources. The Singapore Ministry of Health launched a digital mental wellness portal (*mindline.sg*) in 2020 that provides barrier-free access to curated local mental health resources, clinically validated self-assessment tools, Artificial Intelligence (AI)-enabled digital therapeutic exercises through the Wysa chatbot, and resources for patient navigation to professional services[14]. Building on this digital-first approach, this national cross-platform mental health awareness was launched to improve awareness of mental health symptoms and patient navigation regarding available avenues for residents to seek or receive help. In this study, we perform a structured analysis of a national-level digitally-enabled health promotion campaign, reporting its associated costs, effectiveness in driving awareness, and key lessons learnt for researchers and practitioners of digitally-enabled health promotion in the public health field.

## II. Methodology

### A. Study Design

This study evaluates a national, digitally-enabled health promotion campaign for mental health awareness using social media platforms Tik Tok, Facebook, Instagram and YouTube. This campaign was conducted to raise awareness about mental health-related content to drive awareness of symptoms, conditions, and patient navigation. This includes the mindline.sg platform developed by the Ministry of Health (MOH) Office of Healthcare Transformation (MOHT), which has been appointed as the digital FSTP for mental health in Singapore. This was a three-month long multi-channel campaign from February 2025 to April 2025 using organic content distributed across YouTube, Facebook, Instagram, and TikTok, added distribution with a curated list of partner Key Opinion Leaders (KOLs), as well as paid content amplification in social media using YouTube.

The WHO's Global Strategy on Digital Health (2020) advocates the use of digital tools in extending health promotion[15], with frameworks such as the Health Belief Model and Transtheoretical Model guiding effective campaign architectures[16,17]. The present campaign operationalised these models through an adapted marketing funnel approach (Awareness → Consideration → Conversion). As the designated digital first-stop touch point (FSTP) for mental wellness in our country, mindline.sg was intentionally positioned as the campaign's primary conversion endpoint in this funnel. Social marketing theory provided an additional theoretical underpinning for this campaign, by emphasising the importance of a multimodal, iterative communication model that uses diverse channels and tailored messaging to influence health behaviours[13].

### B. Outcomes and Evaluation

Primary outcome measures were anonymised performance analytics, including engagement indicators of impressions, unique reach, video content views, as well as organic post likes, comments, shares, and saves. These were analysed in a stratified manner to provide information on audience segmentation for public health promotion, including demographic breakdowns (age, gender), device types, and other relevant audience segments (e.g. based on occupation/field of work).

Secondary outcome measures were based on both proximal and distal indicators of campaign impact. Proximal measures included cost-efficiency metrics such as cost per mille (CPM; cost per 1,000 impressions), cost per click (CPC), and cost per action (CPA), while distal impact was assessed through changes in search and traffic to *mindline.sg*. Tools used for analysis include native social media analytics (e.g. Meta insights, TikTok analytics), Metabase which is an internal analytics platform used for monitoring of web traffic, and Google AdWords to evaluate de-duplicated content reach.

### C. Health Promotion Content Development

Each content piece targeted key messages such as symptom awareness, coping strategies, or patient navigation to relevant resources.

The content strategy was informed by a marketing funnel framework comprising three objective of awareness, consideration, and conversion, with content tailored accordingly.

1. Awareness: Narrative videos titled ”Inspector Quek: Dark Place Casefiles” (Chapters 1 to 3).
2. Consideration: Watchlist and Report infographics with symptom highlights and coping strategies and Ask Mili Anything.
3. Engagement/Conversion: Direct Links to mindline.sg, "I Found mindline" testimonials with call-to-action and others.

The long-form content of the campaign centred on a curated narrative video trilogy featuring a fictional "Inspector Quek" character. In these three linked episodes, an investigator recounts mental health case vignettes—specifically Chapter 1: *Anxiety*, Chapter 2: *Unsaid Misery* (hidden depression), and Chapter 3: *Internal Torment*. These professionally produced short films were designed with a distinct "quirky" detective-themed aesthetic to appeal to younger audiences. To reinforce engagement and move viewers further down the funnel, each video ended with a clear on-screen call-to-action ("And you can count on mindline.sg to relieve some pressure with a chat") along with hyperlinks for easy navigation. To amplify reach, the trilogy was supplemented by additional content: infographics on coping strategies, symptom "watchlist" videos, and social media Q&A posts (e.g., "Ask Mili Anything") covering stress management and mental health tips. The content is available at this link: https://www.youtube.com/playlist?list=PLbA8QLV9zYP2ymsWhd7Jy6rZZOh0m3UCk.

### D. Health Promotion Content Dissemination

All content was distributed via major social media channels. The main videos and supporting materials were organically disseminated through cross-posting on YouTube, Facebook/Instagram, and TikTok channels as well as through a curated list of partner social media key opinion leaders (SM-KOLs. Paid amplification was deployed on YouTube channels for long-form video content. For platforms that allowed re-direction of traffic to targeted hyperlinks, we directed users to digital touch points such as *mindline.sg* with avenues for users to connect with community-based mental health service providers. The health promotion campaign was conducted in an anonymised manner to reduce the perceived stigma and barrier to accessing reliable information about mental health.

## III. Results

This campaign generated an overall reach of 3.49 million total impressions, with an estimated unique reach of 1.39 million Singapore residents (33.3% of the total 4.18 million resident population in Singapore). Total video views reached 630,000, with 14,953 likes, 454 shares, 142 comments, and 22 saves. The video on Anxiety had the highest engagement across platforms. The total cost for video content production, dissemination, along with paid amplification in Youtube was SGD $93,796. The cost-efficiency metrics were: CPM of $26.90, CPC of $29.33, and CPA of $6.06. A detailed breakdown of these primary and secondary outcome measures are included in Table 1. Next, we performed audience segmentation analysis for the long-form video content hosted in YouTube with data extracted from paid amplification. Age-segmentation analysis (**Table 1**) showed that the 25–34 age group had the lowest CPC ($1.58) for this mental health content, followed by the 35–44 group ($1.92). In contrast, users aged 45 and above showed higher CPCs, peaking at $4.58 in the 45–54 group. CPMs were generally consistent across age groups (~$1.85–$1.95).

TABLE I. CONTENT REACH AND PERFORMANCE EVALUATION OF CAMPAIGN STRATIFIED BY AGE, GENDER AND DEVICE TYPE

| Segment | Impressions | Views | Clicks | Cost | CPM/ View Rates[a] for "Device" | CPC |
|---|---|---|---|---|---|---|
| **Age Group** | | | | | | |
| **18 – 24** | 527,450 | 34,291 | 501 | $1,024.88 | $1.94 | $2.05 |
| **25 – 34** | 797,228 | 73,395 | 937 | $1,478.08 | $1.85 | $1.58 |
| **35 - 44** | 1,154,847 | 102,948 | 1,168 | $2,241.15 | $1.94 | $1.92 |
| **45 - 54** | 289,680 | 27,929 | 118 | $540.66 | $1.87 | $4.58 |
| **55 - 64** | 29,970 | 3,189 | 12 | $48.41 | $1.62 | $4.03 |
| **65+** | 34,713 | 4,138 | 18 | $58.15 | $1.68 | $3.23 |
| **Unknown** | 648,060 | 51,835 | 443 | $1,204.17 | $1.86 | $2.72 |
| **Gender** | | | | | | |
| **Female** | 1,288,530 | 123,972 | 1,372 | $2,510.99 | $ 1.95 | $1.83 |
| **Male** | 1,604,454 | 123,596 | 1,401 | $2,992.41 | $ 1.87 | $2.14 |
| **Unknown** | 588,964 | 50,157 | 424 | $1,092.12 | $ 1.85 | $2.58 |
| **Device** | | | | | | |
| **Tablets** | 449,927 | 12,645 | 412 | $619.57 | [a]2.81% | $1.38 |
| **TV screens** | 1,317,097 | 202,895 | 244 | $2,773.76 | [a]15.40% | $2.11 |
| **Computers** | 783,551 | 25,655 | 234 | $1,654.11 | [a]3.27% | $2.11 |
| **Mobile phones** | 931,321 | 56,524 | 2,307 | $1,547.94 | [a]6.07% | $1.66 |
| **Campaign Total** | | | | | | |
| **Overall Population** | **3,481,948** | **297,725** | **3,197** | **$6,595.50** | **$1.89** | **$2.06** |

Legend: CPM, Cost per Mille; CPC, Cost per Click.

[a] View Rates (proportion of impressions resulting in a view)

Gender-segmentation analysis (**Table 1**) of reach and performance indicated that female users had the lowest CPC ($1.83) and the greatest cost-efficiency for the reach of this health promotion campaign. Notably, we also found that this campaign resonated most strongly with individuals from the technology and education sectors. In addition, we performed device-segmentation analysis (**Table 1**). Of note, audience members who accessed the content via television screens (TVs) recorded the highest views (202895) as well as view rate (15.4%), while those who accessed it via mobile phones had the greatest click through rates (2307/3197, 72.2%).

Next, we performed stratified analysis of our long-form video content for raising mental health awareness with breakdown by audience affinity to understand audience behaviour. We observed the greatest engagement among viewers from the Technology and Financial industries who had 4.7-fold and 2.4-fold greater representation in impressions, respectively, compared to other audience segments. Additionally, viewers from large employers (250-10,000 employees) demonstrated the strongest engagement, with a 2.2-fold greater representation in impressions relative to other audience segments. Lastly, a substantial increase in search and traffic to mindline.sg was observed during the campaign period, aligning with the campaign's goal of enhancing digitally-enabled patient navigation through Singapore's digital FSTP for mental health (**Figure 1**). This was exemplified by the increase in searches for Mindline-related keywords in the google search engine results page (SERP) after the campaign start in February 2025, indicative of brand re-call.

Fig. 1. Trend in Search and Web Traffic to Mindline.sg

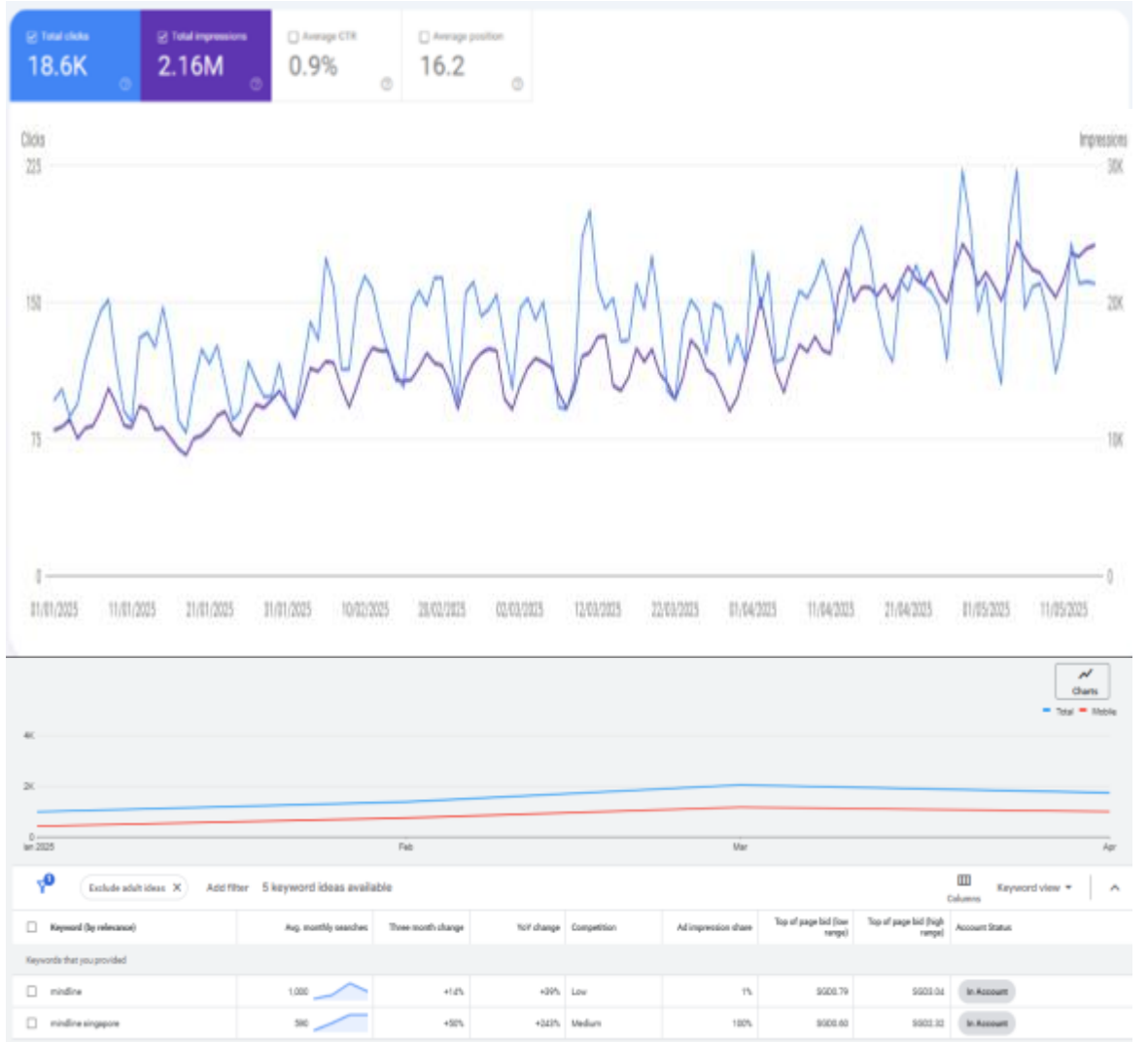

## IV. DISCUSSION

This study illustrates both the promise and limits of contemporary digitally-enabled health promotion campaigns. By leveraging narrative storytelling and multi-platform distribution, the campaign achieved substantial reach and engagement of over 30% of the entire resident population of the country of Singapore (unique reach of 1.39 million Singapore residents out of the total 4.18 million resident population in Singapore). These results align with the goals of digitally-enabled health promotion and social marketing principles: using a multimodal communication strategy (video + infographics + interactive posts) that effectively captured audience attention[7,13]. The "quirky" detective theme and consistent branding helped the campaign stand out in a crowded media landscape, leaving the audience with an impression from the uniqueness associated to this specific campaign[18].

This campaign applied an adapted marketing funnel approach for an initial campaign designed to increase topical recognition about mental health and digital touchpoints for help-seeking, by guiding the audience through successive stages (Awareness → Consideration → Conversion). The content type was optimized for each stage and corresponding performance metrics summarized in **Figure 2**. The awareness phase demonstrated strong brand visibility with unique impressions from over a third of Singapore's population, highlighting the effectiveness of multimodal strategies (narrative videos and infographics) across broad distribution platforms such as YouTube, Instagram, Facebook, and TikTok. At the consideration stage, 630,000 video views and over 15,000 user interactions (likes, shares, comments and saves) suggest that relatable, user-facing formats such as interactive Q&A content and testimonials were effective in deepening engagement and encouraging discourse around mental health. Lastly, testimonials with call-to-actions and direct hyperlinks in the conversion stage lead to 3,000 direct click-throughs to mindline.sg and increased platform traffic during the campaign period, indicating meaningful transition from awareness to action.

Fig. 2. Marketing Funnel Approach and Performance Evaluation of a National-Level, Digitally-Enabled Health Promotion Campaign Trend in Search and Web Traffic to Mindline.sg

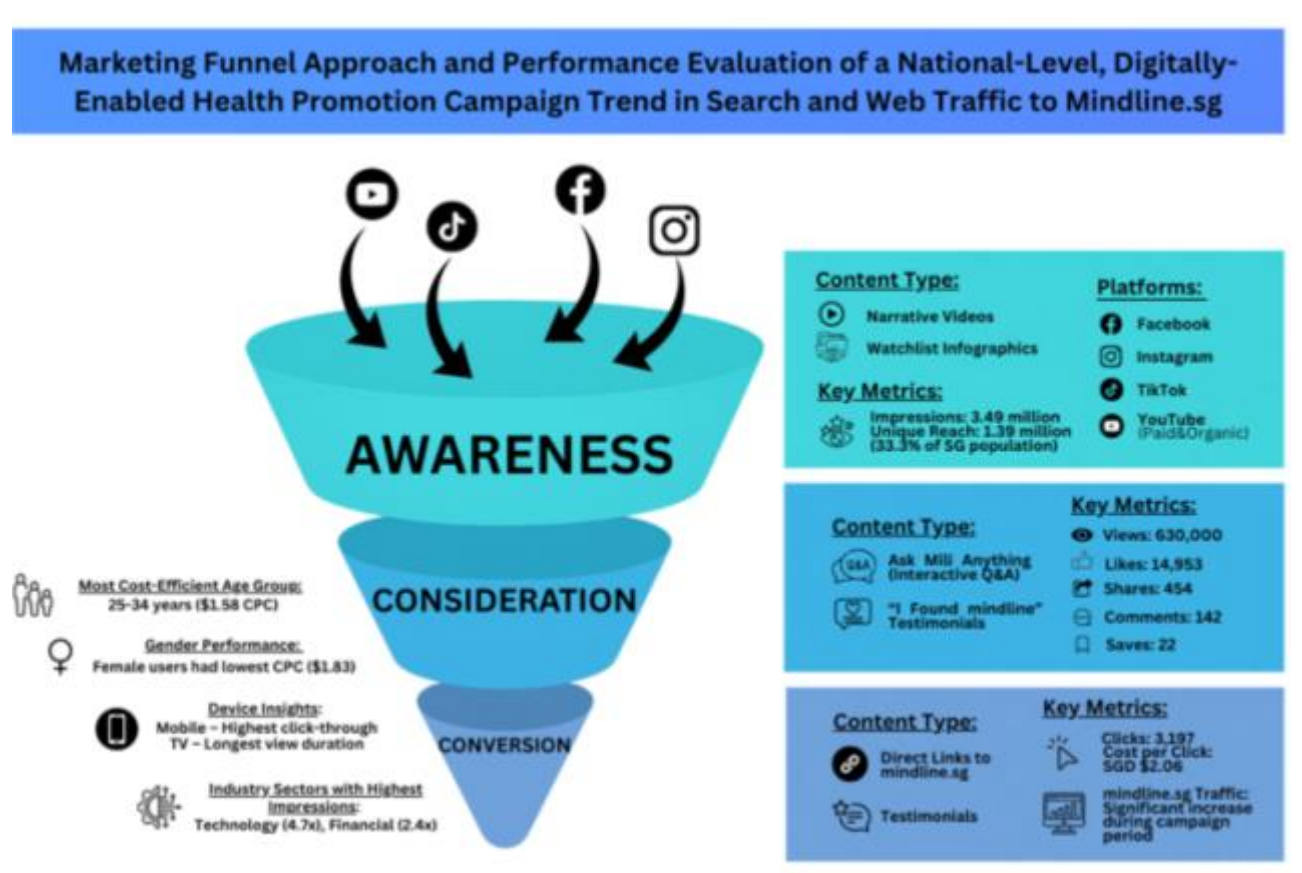

Measured outcomes in the millions for impressions and hundreds of thousands of views in this study demonstrated successful audience penetration when compared with the reports from similar prior work in this field[19]. Frameworks such as the Health Belief Model (HBM), Social Cognitive Theory (SCT), and the COM-B model (Capability, Opportunity, Motivation - Behaviour) have been adapted for similar digital campaigns in the past[20]. Abroms et al previously outlined a set of recommendations specifically for designing social media health campaigns, which focus on the importance of message development, audience segmentation, and performance tracking using digital analytics[20]. Furthermore, Kreuter et al found that narrative-based digital content significantly enhances engagement with public health messaging[21]. Our campaign incorporated these best practice principles through a multimodal content strategy that combined core "interest-building" narratives with practical "coping tips" and interactive avenues embedded in Q&A content.

In using these approaches, we found that a diversified content distribution strategy supported broad awareness and segmental audience engagement. Younger adults (18–44) emerged as the most responsive population, aligning with known social media platform demographic trends: TikTok and Instagram are primarily used by individuals aged 18-29, while Facebook and YouTube are most frequently accessed by those aged 30-49[22]. Moreover, high performance among professionals from the technology and education sectors supports the value of strategic audience segmentation. However, engagement levels among older age groups (>44 years old) and male users were lower, indicating a potential need for differentiated strategies to more effectively engage these segments. These findings emphasize that designing personalised, risk-based messaging based on platform demographic trends may improve content receptivity and potentially increase long-term behavioural impact. Furthermore, optimising content for specific device formats can enhance desired outcomes, for example, increasing completion rates of long-form videos when optimised for television screens, or improving digital conversions through higher click-through rates when content is tailored for mobile devices. Lastly, our relatively low cost-per-view indicates efficient dissemination and supports the viability of such large-scale digital health campaigns.

The findings from this study were consolidated with the strategies and outcomes of other similar digitally-enabled health campaigns conducted in various countries summarised in **Table 2**. While all campaigns leveraged the widespread reach of social media to achieve their respective public health goals, their distinct approaches and unique contextual differences offer valuable insights. Compared to the stroke awareness campaign described in Tunki et al[23], which reached 8.6% of the Nepalese population, the current mental health campaign managed to reach approximately 33.3% of Singapore's resident population. This contrast highlights the importance of digital penetration in enhancing the effectiveness of digital public health campaigns, where the widespread use of digital devices and high rates of digital literacy in Singapore likely contributed to greater engagement. Even in utilising social media, platform choice also plays a critical role as evidenced in the safer sex public health campaign described by Hanson et al[24]. The campaign saw a higher level of engagement with the target age group on Snapchat compared to Instagram (78% vs 45%), highlighting how platform-specific demographics can also affect the effectiveness of digitally-enabled campaigns.

TABLE II. COMPARISON OF OUTCOMES OF DIGITALLY-ENABLED HEALTH PROMOTION CAMPAIGNS

| Study | Country | Purpose | Campaign Design | Duration | Velocity (Impressions/Week) | Content Reach and Performance Evaluation | | | | | | |
|---|---|---|---|---|---|---|---|---|---|---|---|---|
| | | | | | | Total Impressions/Platform(s) | Total Unique Reach | Total Engagements | Cost of Content Production & Dissemination | Cost of Content Amplification | CPM | CPC |
| **Current Study** | SG | To improve awareness of mental health symptoms and patient navigation regarding available pathways for residents to seek | Social media platforms such as YouTube, Facebook, Instagram and TikTok were used to disseminate content pieces targeted at either symptom awareness, coping strategies, or patient navigation to relevant resources from February to April 2025. The content was tailored according to a marketing funnel framework comprising three objectives: awareness, consideration, and conversion. Paid content amplification in social media was performed using Youtube. These materials allowed for the re-direction of traffic to mindline.sg with avenues for users to connect with mental health service providers. | 3 Months | 290,162 | 3,481,948<br>YouTube, Facebook, Instagram, and TikTok | 1,390,000 | 18,768 | SGD $87,200.50 for Long-form video content on mental health education on Youtube | SGD $6,595.50 for YouTube | SGD $26.90 | SGD $29.33 |
| **Hunt et al., 2023[16]** | USA | To raise awareness of the importance and effectiveness of the COVID-19 vaccine to increase vaccination rates | Video and still-image advertisements were disseminated on Facebook as part of a three-phase public health campaign conducted from April to June 2021. The campaign featured racially diverse physicians delivering personalised messages, integrated direct links to multilingual helplines, and involved religious leaders from various communities to raise awareness about the importance of vaccination. | 3 Months | 14,288,029 | 171,456,348<br>Facebook | 25,223,949 | 31,605 | USD $693,176 for Short-form Video and Still-image advertisements on Facebook (Breakdown not specified) | | USD $27.48 | NA |
| **Tunki et al., 2023[23]** | Nepal | To increase stroke awareness among the Nepali population and subsequently improve their understanding of the disease | Between January and June 2022, social media sites such as Facebook, Instagram, Twitter and Tiktok were used to distribute key infographics and images targeted at improving the understanding of stroke. Content was disseminated via both free posts and paid-for promoted posts on these platforms. A website was also created to provide evidence-based and comprehensive stroke knowledge. Content was available in both the English and Nepali languages. | 6 Months | 313,081 | 7,513,952<br>Facebook, Instagram, Twitter, and TikTok | 2,411,505 | 265,095 | EUR $3,474 for marketing specialists, campaign materials, and short videos on Facebook, Instagram, Twitter, and TikTok (Breakdown not specified) | | EUR $0.24 | EUR $0.01 |
| **Hanson et al., 2024[24]** | UK | To promote safer sex, this campaign raises awareness of free condoms and sexual health to registered youths aged 13 to 24 | Between 15 to 21 July 2020, advertisements on Instagram and Snapchat were used to direct individuals who clicked to a Kingston Borough-specific page for registration. | 1 Week | 15,000 | 15,000<br>Instagram and Snapchat | NA | 709 | EUR $140 | EUR $0 | EUR $0.009 | EUR $0.20 |

Legend: CPM, Cost per Mille (cost per 1,000 impressions); CPC, Cost per Click (average cost per user click); SG, Singapore; USA, United States of America; UK, United Kingdom.

Nevertheless, this study has similar limitations to prior work in digitally-enabled health promotion campaigns targeting awareness, whereby tracked outcomes reflect proximal engagement, and do not directly translate to behaviour change. Due to the anonymised nature of data collection to minimise stigma in this campaign, no longitudinal data, survey or referral data were collected that could be used to measure behaviour change. These can be included in future studies to track the impact of digitally-enabled health promotion on behaviour change. They could be incorporated through privacy-preserving approaches in future marketing campaigns designed to drive behaviour change, such as for self-help, reducing stigma or improving help-seeking behaviour, along with additional marketing funnel stages for loyalty, retention, and/or advocacy[25]. Furthermore, self-selection bias may limit the reach of this campaign among disengaged populations, highlighting the importance of omni-channel approaches blending broad online campaigns with targeted offline campaign strategies[26,27]. Moreover, causality remains uncertain for engagement- and awareness-oriented campaigns without control groups or post-campaign behavioural evaluations[28].

Finally, recent findings in the marketing field include reports about the increasingly non-linear and potentially drawn-out nature of customer journeys today, with audience members encountering multiple touchpoints across digital channels, and contemplating conversion over months[29]. The use of behavioural re-targeting in digital platforms can help address duplicate accounts that users may maintain across various platforms, to show relevant content with retargeted messaging to optimize influence and impact of DEHP[30]. This also emphasises the need for longitudinal brand-building and data collection on intent, along with a potential role of artificial intelligence (AI). AI techniques such as machine learning (ML) could be applied in future campaigns to realise the value from audience segmentation and brand advocates across multi-channel marketing for the prediction of intent.

However, strengths of this study include its thematic coherence, creative storytelling, and multi-platform reach. The use of audience segmentation analysis across social media platforms provided valuable insights into the potential reach and resonance of digitally-enabled health promotion efforts. Furthermore, device-specific analytics revealed how user behaviour differs across medium, such as a greater depth of consumption of video-based content with content accessed through television devices and greater engagement with click-throughs with content accessed through mobile devices. These insights will be useful in guiding the deployment and design of future online public health campaigns, which could potentially explore the use of online health communities (OHCs) for baseline/comparison groups[31], and post-campaign surveys to establish causality of digitally-enabled health promotion campaigns on behaviour change[32,33].

## Conclusion

This national-level mental health campaign achieved compelling reach and engagement through a creative, narrative-driven strategy. It demonstrates the potential scalability of digitally-enabled health promotion to reach large audiences. Future campaigns could integrate digital and physical outreach while incorporating robust post-campaign evaluation strategies to further understand behaviour change.